# Time-dependence of SrVO$_3$ thermionic electron emission properties

*Md Sariful Sheikh[1], Ryan Jacobs[1], Dane Morgan[1,*], John Booske[2,*]*

[1] Department of Materials Science and Engineering University of Wisconsin-Madison, Madison, WI, 53706, USA.

[2] Department of Electrical Engineering and Computer Science University of Wisconsin-Madison, Madison, WI, 53706, USA

## 1. Abstract:

Thermionic electron emission cathodes are critical components of various high power and high frequency vacuum electronic devices, electron microscopes, e-beam lithographic devices, and thermionic energy converters, which all demand an efficient and long-lasting low work function cathode. Single phase, polycrystalline perovskite oxide SrVO$_3$, with its intrinsic low effective work function and facile synthesis process, is a promising cathode candidate, where previous works have shown evidence of an effective work function as low as 2.3 eV. However, assessment of the stability over time under conditions relevant for operation and the related interplay of evolving surface chemistry with emission performance are still missing, and necessary for understanding how to best prepare, process and operate SrVO$_3$ cathodes. In this work, we study the vacuum activation process of SrVO$_3$ and find it has promising emission stability over 15 days of continuous high temperature operation. We find that SrVO$_3$ shows surface Sr and O segregation during operation, which we hypothesize is needed to create a positive surface dipole, leading to low effective work function. Emission repeatability from cyclic heating and cooling suggests the promising stability of the low effective work function surface, and additional observations of drift-free emission during one hour of continuous emission testing at high temperature further demonstrates its excellent performance stability.

## 2. Introduction:

Thermionic electron emission cathodes are an essential component of various types of travelling wave tubes, magnetrons and other vacuum electronic devices (VEDs) which are widely used in satellite communications, RADAR systems, remote sensing, medical imaging, security, high-energy linear accelerators, microwave ovens, and fluorescent lamps [1-6]. The high emission current density and ability to provide long-term stable emission make thermionic cathodes highly popular for the reliable operation of various high power and high frequency VEDs [7, 8]. Besides high power VEDs, thermionic cathodes are also widely used in lower power devices such as electron microscopes, e-beam lithography, X-ray generation, materials characterization, and have prospective applications in atomic manipulation (e.g., precise positioning of dopant atoms in crystal lattices using focused electron beam), thermionic conversion of concentrated sunlight and industrial waste heat into electricity, and ion-thrusters for deep space exploration [9-14].

The emission current from a cathode surface is largely determined by its work function, which is defined as the minimum energy required to expel an electron from the cathode surface to vacuum. The work function is a highly local quantity, and essentially all thermionic emission cathodes feature a complex, heterogeneous emitting surface with multiple different crystallographic orientations and surfaces. Such a complex emitting surface results in a distribution of work function values, which, depending on the characterization approach, may mix to form a so-called "effective" work function. When characterizing the performance of thermionic cathodes, it is common practice to quote a single effective work function value as a metric of the cathode performance, for example, by measuring the emission current versus temperature and finding the corresponding Richardson-Dushman emission curve (and corresponding effective work function) which most closely corresponds to the thermionic emission data. A detailed review of work function physics, measurement, characterization and applications is given in the review by Lin et al.[15] A lower work function is desirable to maximize thermionic emission current, hence, various strategies have been developed to decrease the cathode surface effective work function over the past century, thereby improving the emission current density, lowering the operating temperature, and enhancing the long-term operational stability of thermionic cathodes [2, 16-19].

Numerous types of thermionic emission cathode materials have been investigated over the past several decades. Single crystal hexaborides, like $LaB_6$ and $CeB_6$, which are extensively used in electron microscopes, electron beam lithography, electron beam welding, X-ray tubes, and free electron lasers, have an effective work function of about 2.7 eV [20]. The 2.7 eV effective work function necessitates a high operating temperature of 1400 – 1550 °C to obtain the necessary thermionic current density [21, 22]. Such high operating temperature results in significant sublimation of the emitting surface, restricting the lifetime of hexaboride cathodes to typically only a few thousand hours [23], resulting in frequent and expensive cathode replacement. Dispenser cathodes dominate the high power and high frequency VEDs market due to their very low effective work function (approximately 1.8-2.1 eV) and higher current density at lower temperature as compared to the early-stage conventional metal (W and Ni), oxide-coated cathodes (Ba-Sr-O) and hexaboride cathodes [2]. A B-type dispenser cathode consists of a porous W-pellet and the pores are filled with a mixture of barium oxide, calcium oxide and aluminum oxide. At operating temperature, Ba diffuses to the W surface and makes a Ba-O dipole layer structure which decreases the effective work function of W cathode from 4.6 to about 2.1 eV [24]. In an M-type dispenser cathode, a film of noble refractory metals such as Os, Ir, Ru, or Re is deposited on the impregnated cathode surface, which further reduces its effective work function to about 1.8 eV, resulting in lower operating temperature and enhanced cathode lifetime [25]. Despite the significant development of dispenser cathodes, they still face long-term operational stability challenges due to the fundamentally volatile nature of dispensing Ba.

Due to the shortcomings of current dispenser and hexaboride cathodes, discovery and development of a single-phase material that exhibits intrinsic low work function comparable to the dispenser cathodes or hexaborides and has robust operational stability would be a major advance for technologies using thermionic cathodes. $ABO_3$-type perovskite oxides, with their easily tunable physical properties could be the ideal class of materials to search for the desired conducting and low work function cathode. $ABO_3$ perovskites consist of alternating layers of the polar AO and $BO_2$ units along the [001] direction, which may provide an intrinsically low work function originating from AO surface dipoles. The work of Jacobs et al. used density functional theory (DFT) calculations and first suggested that $SrVO_3$ is a promising material, where its SrO terminated (001) surface is predicted to have a low 1.9 eV work function [26]. Later, Ma et al. performed a high-throughput DFT screening study of more than 2000 perovskite oxide materials [27]. According to their calculations, conducting and thermodynamically stable perovskite oxides, such as $BaMoO_3$, $SrMoO_3$, and Nb-doped $SrTiO_3$ may offer < 2 eV work function on their AO-terminated (001) oriented surface. In order to experimentally verify the DFT predicted results, Lin et al. measured the effective work function of perovskite oxide $SrVO_3$ and $BaMoO_3$ cathodes by measuring thermionic emission current density and observed generally reproducible effective work functions of about 2.7 eV and 2.65 eV for $SrVO_3$ and $BaMoO_3$, respectively [28, 29]. Besides these modestly low effective work functions of about 2.7 eV, polycrystalline $SrVO_3$ cathodes have shown several instances of very low effective work function of about 2.3 eV [28, 30]. This effective work function of 2.3 eV is, to our knowledge, the lowest experimentally observed effective work function reported from any bulk, polycrystalline, conductive oxide material without any doping or volatile surface coating. If realized in a reproducible manner, this intrinsic low effective work function cathode could significantly ease the cathode fabrication process, as it does not require any special surface coating or doping, and may provide more reliable operation as compared to volatile Ba-containing dispenser cathodes.

In this work, we expand on the work of Lin et al. with more detailed study of the time-dependent performance of bulk, polycrystalline $SrVO_3$ thermionic cathodes. We studied the vacuum activation process and the resulting cathode microstructure and surface chemistry evolution, the operational stability at high temperature and in ultra-high vacuum over a period of 15 days, and the short-term operational stability using a continuous emission test at operating conditions. This work provides practical stability assessments and understanding of the evolution and interplay of surface chemistry with measured effective work function of $SrVO_3$. This information is important for realizing the practical application of $SrVO_3$ as a thermionic emission cathode material and can provide a future basis for improved preparation and processing of $SrVO_3$ cathodes to realize exceptional emission behavior.

### 3. Experimental section:

#### 3.1. SrVO₃ synthesis:

The $SrVO_3$ perovskite oxide powder was prepared using the sol-gel synthesis method [28]. In order to synthesize 1 gram of $SrVO_3$ powder, 0.898 grams $Sr(NO_3)_2$ (Sigma Aldrich, 99.9%), 0.712 grams $NH_4VO_3$ (Sigma Aldrich, 99%) and 25 grams citric acid (Dot Scientific Inc.) were dissolved in 250 ml deionized water (minimum resistance 18 MΩ-cm) in a 500 ml alumina beaker using a magnetic stirrer. The homogenous solution was dried on a hot plate (~ 80 °C) until it evaporates most of the water and forms a gel. The alumina beaker, along with the gel, was transferred into a box furnace (ThermoFisher Scientific, Model: BF51866C-1) and heated at 120 °C for 12 hours to remove all of the water. The furnace temperature was raised to 600 °C at a heating rate of 2 °C/min and sintered in air at 600 °C for 12 hours to remove all other organic components. The dried off-white precursor powder was ground for 30 min using an alumina mortar and pestle. Finally, the ground precursor was calcined at 1050 °C for 10 hours in a reducing environment using a tube furnace (ThermoFisher Scientific, Model: STF55433C-1) of tube dimension OD ~ 5.1 cm, ID ~ 4.5 mm, L ~ 121 cm (Material > 99.6 % pure alumina, Advalue Technology). The reducing environment was maintained by flowing 5 % $H_2$ / 95 % Ar gas at a flow rate of 200 SCCM throughout the calcination process.

Figure S1(a) in the supporting information (SI) shows the X-ray diffraction (XRD) pattern and corresponding Rietveld refinement of the synthesized $SrVO_3$ powder, confirming the phase pure synthesis of the perovskite oxide. The Rietveld refinement reveals its cubic perovskite structure ($Pm\bar{3}m$) and the obtained lattice parameters are represented in Table S1 in the SI file. The refined lattice parameters a = b = c = 3.847 (8) Å are consistent with previous report, a = b = c = 3.842 Å (PDF: 04-006-0739).

### 3.2. Cathode preparation:

To prepare the cathode, the synthesized $SrVO_3$ powder was pressed into a cylindrical pellet (diameter ~ 3 mm and thickness ~ 0.75 mm) using 300 MPa pressure at room temperature. The compressed pellet was placed on an alumina boat and sintered at 1350 °C for 10 hours in a reducing environment using the same tube furnace from the calcination step above. During the pellet sintering process, first, the temperature was raised from room temperature to 1350 °C at a heating rate of 4.5 °C/min and sintered at 1350 °C for 9 hours. During this entire heating up and sintering process, a constant partial hydrogen pressure of 0.1 % was maintained by flowing 4 SCCM of 5% $H_2$/95% $N_2$ gas and 197 SCCM of pure Ar gas mixture. At the end of this step, the partial hydrogen pressure inside the tube was increased to 5% by flowing a 5% $H_2$/95 % $N_2$ gas mixture at a flow rate of 200 SCCM, and the sample was further sintered at 1350 °C for 1 hour to intentionally over-reduce the top surface. After a total of 10 hours of high temperature sintering at 1350 °C, the sample was cooled down to room temperature at a cooling rate of 4.5 °C/min. The obtained pellet is referred to as the *"as-sintered cathode"* in the rest of the manuscript. The as-sintered $SrVO_3$ cathode surface was over-reduced, with the presence of some secondary non-perovskite phases such as $Sr_2VO_4$, $V_{1-\delta}O$, $Sr_3V_2O_{7-\delta}$ as shown in Figure S1(b) in the SI. The $SrVO_3$ cathode was prepared after removing this over-reduced surface. The over-reduced top surface of the as-sintered cathode was removed by manually scraping using a sharp knife edge, where the perovskite $SrVO_3$ phase was exposed on the top of the sample. The electron emission tests were performed on this scraped pellet, which is referred to as the *"pre-emission cathode"* in the rest of manuscript. We studied the microstructural and surface chemical properties on the cathode surface before emission test and after electron emission test. The cathode after electron emission test will be referred to as the *"post-emission cathode"* in the rest of the manuscript.

### 3.3. Characterizations:

The XRD pattern of the perovskite oxide powder, as-sintered cathode, pre-emission cathode and post-emission cathode were studied using a Cu-Kα X-ray diffractometer (Bruker D8 Discovery). The composition and microstructure evolution of the perovskite oxide cathodes was examined using a field effect scanning electron microscope (FESEM, Zeiss 1530). Surface elemental stoichiometry of the pre-emission cathode and post-emission cathode was analyzed using energy dispersive spectroscopy (EDS) attached with a Zeiss 1530 FESEM, and X-ray photoemission spectroscopy (XPS) Thermo Al-Kα X-ray photoelectron spectrometer. The atomic percentages on the cathode surfaces were calculated after performing Shirley-type background corrections using an integrated data processing software of the Thermo Al-Kα XPS instrument. To study the bulk stoichiometry of the pre-emission cathode and post-emission cathode, we used $Ar^+$ ion sputter etching under high vacuum in the XPS chamber to remove the top surface and performed elemental analysis with XPS in the exposed surface. We then used further $Ar^+$ ion sputter etching to reach deeper layers in the post-emission cathode and studied those with XPS. Monoatomic $Ar^+$ ions with a kinetic energy of 4000 eV at an angle of 60° from the sample surface were sputtered on the sample. We repeated the $Ar^+$ ion etching several times and calculated the elemental ratio at different depths below the surface. Using this similar sputtering condition, Bourlier et al, observed an etching rate of 0.03 nm/s in their $SrVO_3$ thin film [31]. Hence, we assumed 1000 seconds of Ar$^+$ ion sputtering on our sample may remove around 30 nm of SrVO$_3$ from the cathode surface.

### 3.4. Electron emission test:

We have performed the electron emission test using a custom in-house laboratory setup [28]. A schematic of the emission test set up is shown in Figure 1. The assembly mainly consists of a W-filament exposed halogen bulb (A: G4 Bi-Pin Base, 12V, 20-Watt, Size T3, life span approximately up to 2000 hours, brand-Taiyaloo, available at Amazon.com market) kept floating at a negative high voltage (approximately 1.2 kV), Mo-sample holder, Mo-foil heat shield, Mo-mask, and anode fixture. Part of the halogen bulb glass case was cut using a diamond blade and the sample holder cap connected to the ground was placed on top of the half-cut halogen bulb as shown in Figure 1. In this study, we used a Mo-sample holder, which indirectly heated up the cathode to the desired temperature. At high temperature, the W filament, which is kept biased at negative high voltage, emits electrons. The emitted electrons hit the bottom of the Mo-sample holder and heat it up along with the cathode pellet (part 'e' in Figure 1) placed on it. The anode fixture consists of grid (a piece of monel-400 mesh with 0.28 mm wire diameter and 0.56 mm$^2$ opening) and catcher (Mo-foil with a hole at the center). The gap between the cathode and grid was approximately 1 mm, and the gap between grid mesh and catcher foil was approximately 6.5 mm. The hot cathode surface is viewable through the grid mesh, catcher hole and glass window as shown in Figure 1. The temperature of the cathode was monitored using a disappearing filament pyrometer (Leeds & Northrup, CO) equipped with a 655 nm red light filter. To calculate the actual temperature (T) from the brightness temperature (T$_b$) of our sample, we assumed the emissivity of our sample to be 1 and determined the effective emissivity after considering the glass window (transmissivity approximately 0.76) and reflection mirror (reflectivity approximately 0.93) effect to be ε = 0.7068. Finally, the actual sample surface temperature was estimated using Planck's law:

$$T = \frac{h\vartheta}{k_B} \frac{1}{\ln\left[1+\epsilon\left(\exp\left(\frac{h\vartheta}{k_B T_b}\right)-1\right)\right]} \qquad (1)$$

The chamber pressure was maintained using a cryogenic pump and an ion pump attached to the sample test chamber. After sample loading, both the pumps were used to reach the desired vacuum < 10$^{-9}$ torr before turning on the emission heater. The cathode temperature was slowly raised by gradually increasing the emission heater filament current while maintaining the chamber pressure below < 1 x 10$^{-9}$ torr. The chamber pressure during emission was in the range of 1x10$^{-9}$ to 5x10$^{-9}$ torr.

To measure the emission current, a constant positive DC voltage of 2.1 kV and a varying 0 to 1.9 kV square wave pulse of frequency 10 Hz and 25 µs-width were applied to the catcher (part 'b' in Figure 1) and grid (part 'c' in Figure 1), respectively. The electron emission current from the SrVO$_3$ cathode was measured as the total current flowing through the grid and the catcher circuit as shown in Figure 1. The current was measured by a current monitor (Pearson Electronics, INC. Model: 4100C) and the output from the monitor was recorded with a LeCroy WaveSurfer 44Xs 400 MHz oscilloscope, where both instruments have an accuracy of ± 1 %. The emitted current density was determined by dividing the measured emission current by the area of the 1 mm diameter mask (part 'd' on Figure 1) placed on the sample (part 'e' on Figure 1). During the vacuum activation process, the sample was kept at the operating conditions of high temperature and ultra-high vacuum for several days, and the emission current was measured after applying the anode voltage at certain time intervals. Short-term continuous emission test was performed by continuously applying the anode voltages for 1 hour and simultaneously recording the emission current from the cathode. During temperature-dependent current density measurement, cathode temperature was varied by varying the emission heater filament current controlled by a variable AC voltage ($V_{ac}$).

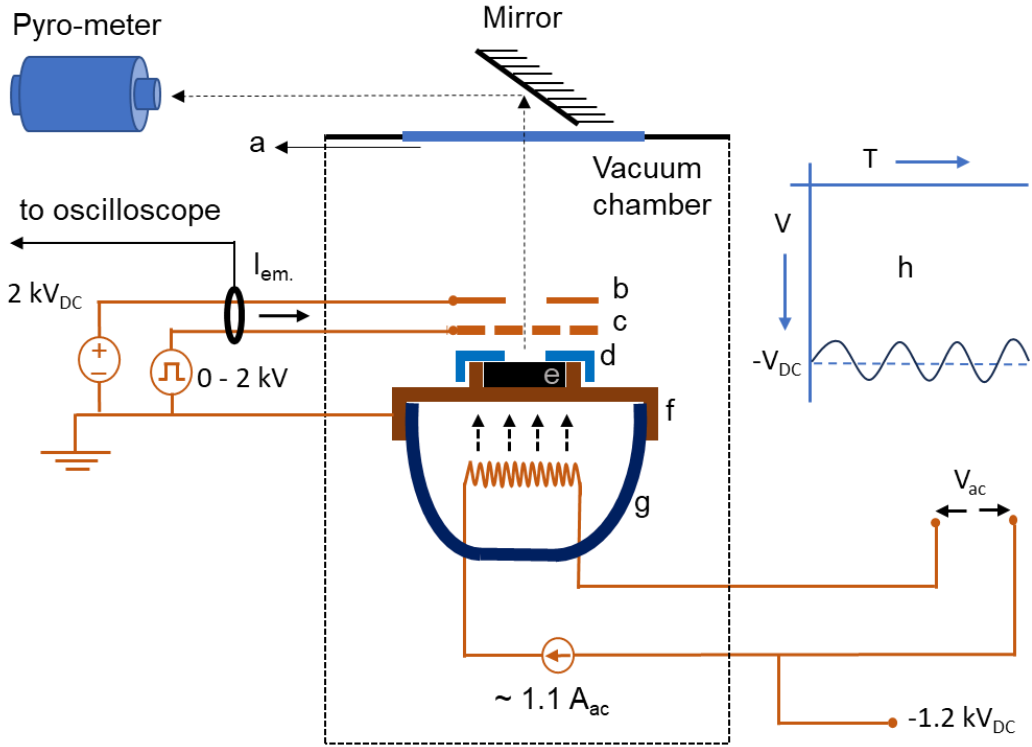

**Figure 1**: A schematic of the electron emission test set up. a, b, c, d, e, f, g and h represent glass window, catcher, grid, Mo-mask with 1 mm diameter hole at the center, SrVO$_3$ cathode, Mo-sample holder cap, half-cut halogen bulb, and voltage profile at the electron emission heater filament, respectively.

## 4. Results and discussion:

### 4.1. Vacuum activation of SrVO$_3$ cathode

We successfully performed long-term vacuum annealing of four scraped as-sintered SrVO$_3$ cathodes (i.e., pre-emission cathodes) at elevated temperature and at vacuum chamber pressure of approximately 10$^{-9}$ torr. We observed gradually increasing electron emission from two cathodes, which were annealed in vacuum chamber at the average temperature of 1260 °C and 1280 °C, and are denoted as cathode A and cathode B, respectively. We observed two other cathodes didn't show the gradually increasing electron emission and their surface stoichiometry are reported in Section 4.3 along with the Cathode A and B. Both cathodes A and B took about 100 hours to show significant electron emission. We speculate that during this high temperature vacuum activation time, the surface damage caused by the removal of the over-reduced phases is recovered, any surface contamination which occurred during air exposure is removed, and, generally, the surface chemistry and structure evolve in a manner which facilitates electron emission (i.e., the SrVO$_3$ cathode is "activated"). Based on our XPS results (discussed in Section 4.3), we believe this activation involves Sr enrichment on the surface, which may be a mechanism to create the desired positive surface dipole (discussed in Sec. 2) and reduce the effective work function. There is evidence for such changes in the evolution of cathode surface chemistry, which is discussed in more detail in Section 4.4 and 4.5. Interestingly, the SrVO$_3$ cathode shows a gradual incremental increase in current density during its vacuum annealing process until it reaches nearly steady emission after around 325 hours, as shown in Figure 2. Emission tests longer than about 300-400 hours were not performed as the halogen bulb filament, part g in Figure 1, broke in both cases where we attempted longer tests. Our samples show continual improvement in emission current density over time, with no evidence of emission decrease, at least during the 400-hour time interval evaluated here. Overall, we have no evidence to suggest that the performance of SrVO$_3$ degrades with time when held at ultra-high vacuum and high temperature, and, to the contrary, the time evolution of SrVO$_3$ emission demonstrates the material activates and becomes better conditioned under operating conditions.

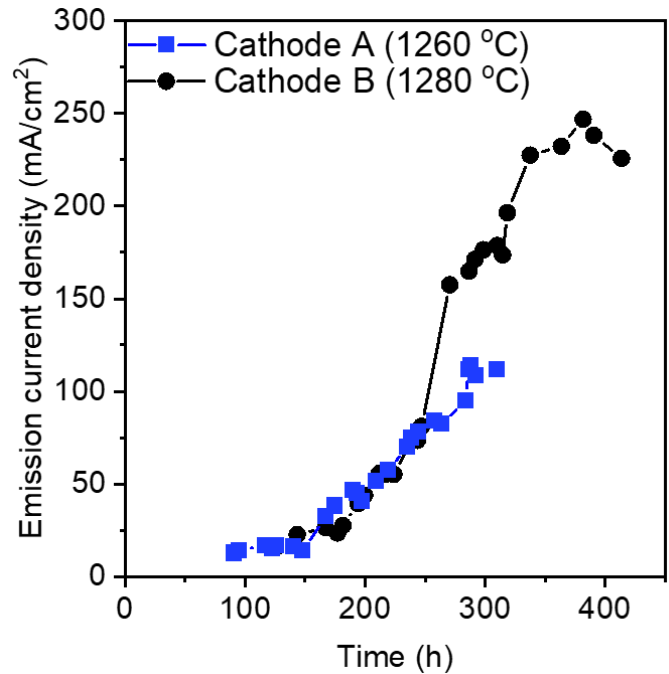

**Figure 2:** Vacuum activation of SrVO$_3$ cathodes. The current density is measured in the temperature-limited regime.

## 4.2. Operational emission stability of activated SrVO$_3$ cathode

We studied the cathode emission performance when the cathode emission current density nearly saturates, and the cathode surface evolution is expected to reach a stable state (e.g., after 325 hours for cathode B). To assess the stability of emission performance from SrVO$_3$ cathodes under thermal cycling, we performed repeated cyclic heating and cooling tests and studied the emission, as shown in Figure 3. The cathode shows similar emission during each temperature cycle, with average ranging over about 22.5±9.4 % from the mean of all tests at a given temperature and with effective work function within about 2.7 - 2.8 eV for all samples in the temperature range from 1050 – 1250 °C. The repeatability of the emission current density during cycling heating and cooling cycle suggests at least short-term thermal stability of the low effective work function surface activated at high temperature and high vacuum. We also performed a short-term continuous pulse electron emission test for one hour at operating temperatures of 1135, 1155, 1205, 1245, and 1260 °C as shown in Figure 4, and found drift-free emission current within this temperature range. Overall, these tests suggest good stability of the SrVO$_3$ cathode and its emission performance when repeatedly heated and cooled and when held at constant emission for a period of 1 hour, a promising initial result for possible commercial use of this cathode.

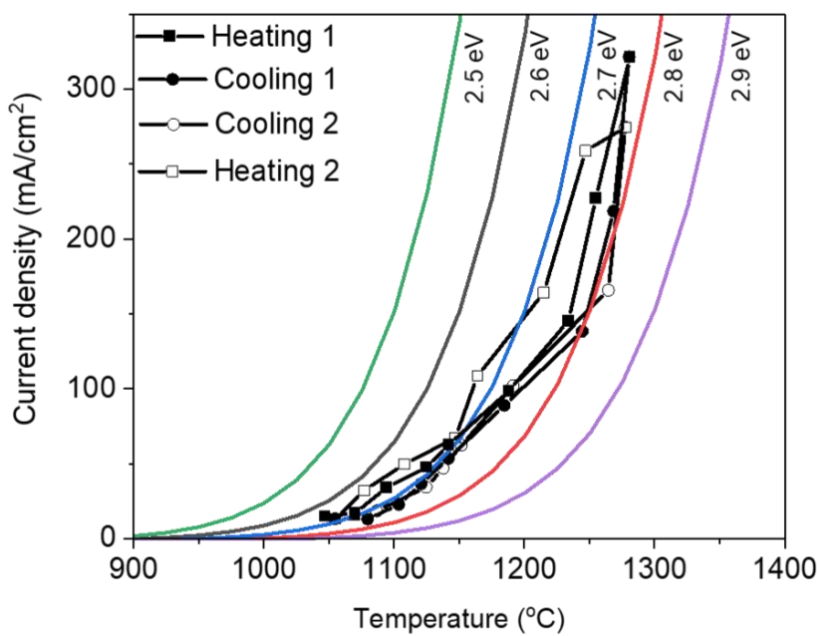

**Figure 3:** Emission current stability test during cyclic heating and cooling process. Emission current was measured at an anode-cathode voltage of 1.9 kV in the temperature-limited regime.

The solid lines represent the theoretical emission current density obtained from the Richardson-Dushman equation for the work function of 2.5, 2.6, 2.7, 2.8 and 2.9 eV.

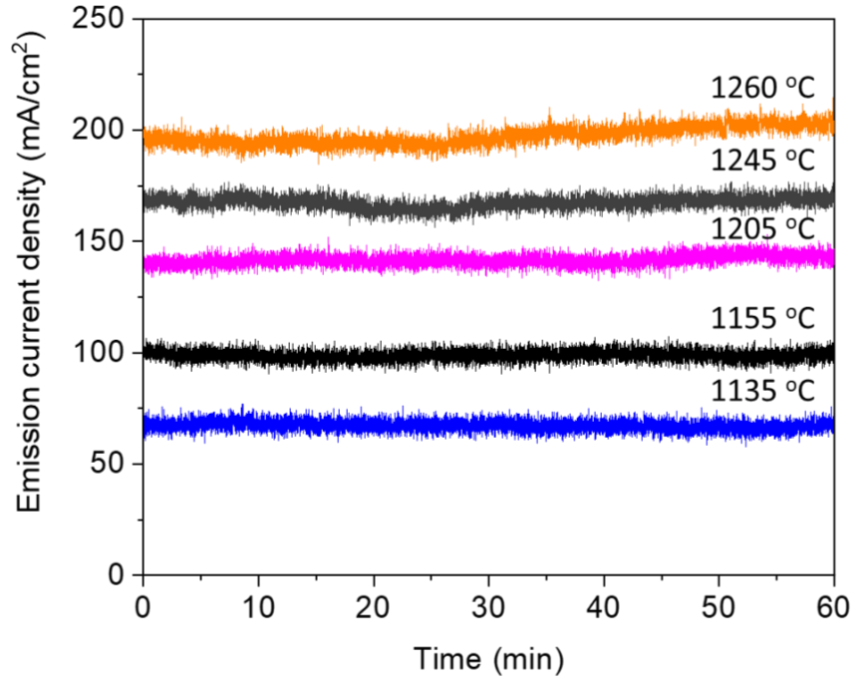

**Figure 4:** Short term continuous pulse emission current stability test of SrVO$_3$ cathode at different temperatures.

### 4.3. Surface stoichiometry evolution during cathode activation

To study the surface evolution over the course of cathode activation and to understand the interplay of the surface chemistry with measured effective work function, we performed X-ray photo emission spectroscopy (XPS) on the pre-emission cathodes and post-emission cathodes. First, we studied the surface stoichiometry evolution during the vacuum activation process of the pre-emission cathodes. We also performed Ar$^+$ ion sputter etching on the post-emission cathode surface to study the bulk stoichiometry of the cathode. The elemental ratio on the pre-emission cathode surface and elemental ratio as a function of the Ar$^+$ sputter time on the post-emission cathode is summarized in Figure 5. The pre-emission cathode showed slight Sr-depletion (Sr/V ratio ~ 0.93 (ideal ratio is 1)) and slight O-depletion (O/(Sr+V) ratio ~ 1.46 (ideal ratio is 1.5)), the latter perhaps due to sintering under reducing gas. In contrast, the post-emission cathode showed significant Sr- and O-enrichment on the surface with much higher Sr/V and O/(Sr+V) ratio than the pre-emission cathode, suggesting that Sr and O migrate to the surface at high temperature and may form some kind of Sr- and O-rich surface termination or secondary phase on the emitting surface. With the present analysis, we cannot tell if any part of this Sr-O-rich surface is the same SrO (001) low work function surface predicted from DFT studies,[26] but our overall finding that having Sr and O on the emitting surface leads to a reduced effective work function and improved emission is consistent with this hypothesis. It is worth noting that similar surface Sr-segregation phenomena has been reported in other perovskite oxides such as $La_{1-x}Sr_xMnO_3$, $La_{1-x}Sr_xCo_{1-y}Fe_yO_3$ and $La_{1-x}Sr_xCoO_3$ [32-34], which show Sr segregation during their use as a solid oxide fuel cell electrode material at temperatures of approximately 800 °C and atmospheric oxygen pressure. We found that the bulk of the post-emission cathode is slightly Sr and O depleted, which might happen due to the Sr segregation to the surface and perhaps also gradual evaporation of Sr from the surface. As the top surface is Sr and O rich, the evaporation rate of V may be low as compared to Sr and O, making the overall cathode effectively V-rich and slightly over reduced as observed in the XRD study in Section 4.6. Despite the potential for some amount of Sr loss, we again note that we have not observed any notable decrease in emission from $SrVO_3$ over the course of 400 hours of operation. We also studied the elemental ratio of one cathode sample which didn't show any emission during the vacuum activation process and found a much reduced Sr/V ratio on its surface (before $Ar^+$-sputtering: Sr/V approximately 0.68) and inside the bulk (after $Ar^+$-sputtering: Sr/V approximately 0.39), consistent with the hypothesis that surface Sr and O segregation is essential to reach the low work function and high emission. More specifically, this finding is consistent with the predictions from DFT that a SrO-terminated surface is needed to realize a low work function (although, as noted above we have no direct evidence that the SrO-terminated surface is present in our high performing samples), and the qualitative idea that a Sr-O-rich surface will create an electropositive dipole that can lead to a low work function.

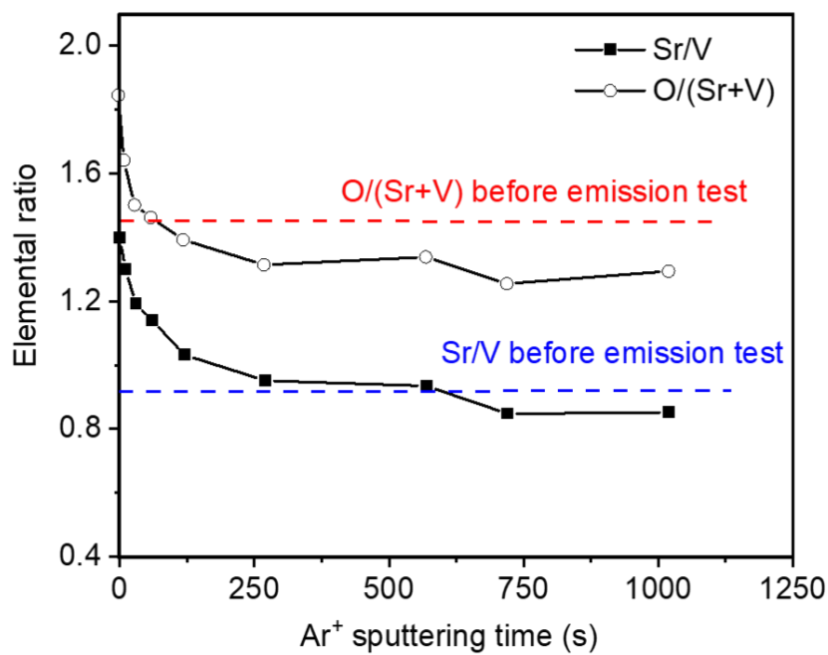

**Figure 5:** Elemental ratio of the cathodes. The red and blue dashed lines indicate the O/(Sr+V) and Sr/V ratios in the pre-emission cathode, respectively. The open circles and filled black squares represent the O/(Sr+V) and Sr/V ratios in the post-emission cathode, respectively.

To study the chemical nature of the emission surface of the cathodes, we examined the core-level Sr-3d, V-2p and O-1s atomic spectra of the pre-emission cathode and post-emission cathode before and after $Ar^+$ sputtering, and fit those spectra using Shirley-type background at CasaXPS software as shown in Figure 6 and Figure S2. A summary of the fitted parameters is given in Table S2 in the SI file. Reconstruction of the Sr 3d peaks results in two chemical states of Sr. The Sr 3d peaks with smaller FWHM are associated with the lattice Sr and the Sr 3d peaks with higher FWHM are associated with the surface Sr-O and hydroxide components [35]. During fitting Sr-3d spectra, a doublet splitting orbital energy gap of 1.75 eV and an aerial ratio of 2/3 were maintained between its $3d_{5/2}$ and $3d_{3/2}$ components. The percentage of the surface Sr in the Sr 3d spectra of the pre-emission cathode is around 9.8 % of total Sr. On the other hand, the post-emission cathode (before $Ar^+$ sputtering) shows significant Sr on the surface (57.7 % of total Sr), suggesting surface Sr segregation during the vacuum activation process of $SrVO_3$ cathode. The surface Sr content is reduced to 18.6 % in the post-emission cathode after $Ar^+$ sputtering, suggesting that Sr segregation is limited to surface regions only. The lower binding energy shifting of Sr-3d and O-1s peaks in the post-emission cathode before $Ar^+$ sputtering as shown in Table S2 suggests that surface of the post-emission cathode has more Sr and O as compared to pre-emission cathode and Ar[+] sputtered post-emission cathode, which is also consistent with the observation of surface Sr and O enrichment during vacuum activation, as observed above.

Overall, our XPS analysis reveals that surface Sr-segregation happens in SrVO$_3$ cathode during its high temperature operation and Sr-atom evaporation from the Sr-enriched surface causes Sr-deficiency within the cathode. However, more research is required to fully understand the Sr-segregation mechanism and to optimize this perovskite oxide cathode for its practical applications.

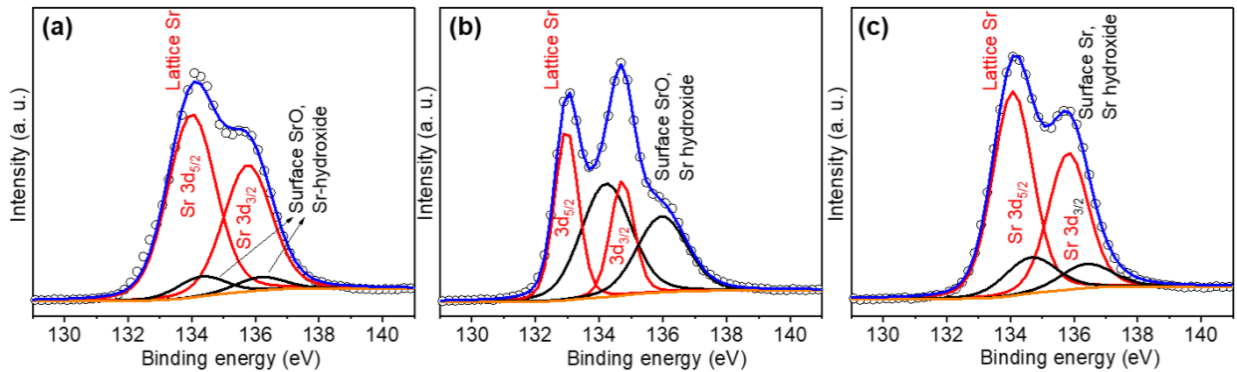

**Figure 6:** Core level Sr-3d atomic spectra of SVO (a) pre-emission cathode (scraped as-sintered pellet), (b) post-emission cathode before Ar[+] ion sputtering, and (c) post-emission cathode after Ar[+] sputtering. Black circles and solid lines represent the experimental and fitted data, respectively.

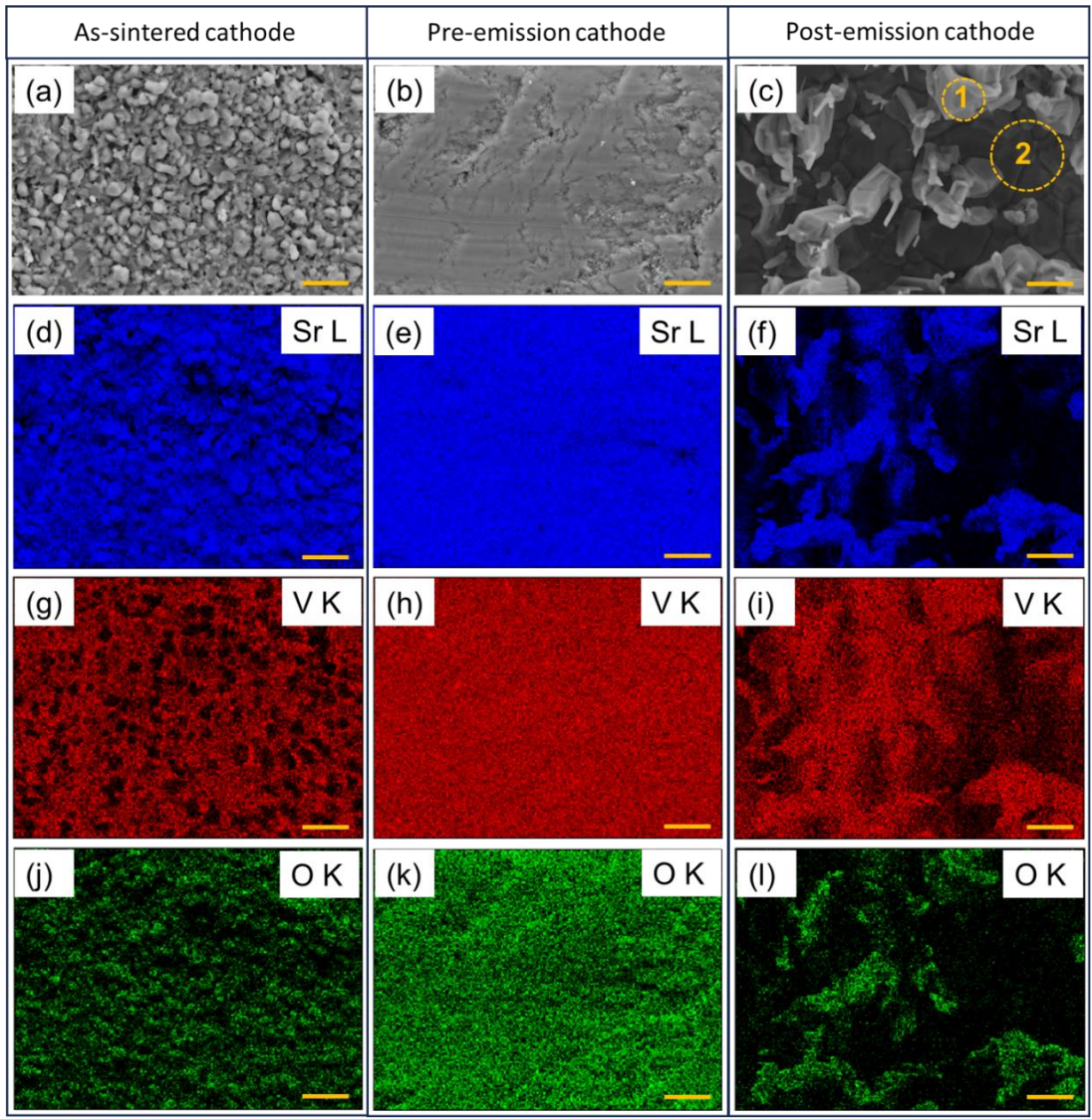

**Figure 7:** FESEM image of (a) as-sintered cathode, (b) pre-emission cathode (scraped as-sintered pellet), (c) post-emission cathode. Sr EDS elemental mapping of (d) as-sintered cathode, (e) pre-emission cathode, (f) post-emission cathode. V EDS elemental mapping of (g) as-sintered cathode, (h) pre-emission cathode, (i) post-emission cathode. O EDS elemental mapping of (j) as-sintered cathode, (k) pre-emission cathode, (l) post-emission cathode. Scale bars in all Figures are 10 μm.

### 4.4. Surface micro-structure evolution during cathode activation

FESEM image of the sintered SVO pellet surface as shown in Figure 7(a) reveals its rough microstructure with 0.5 to 2 μm grain size. Figure 7(b) shows the surface morphology of the pre-emission cathode (scraped as-sintered pellet). The surface morphology of the post-emission cathode as shown in Figure 7(c) reveals a significant microstructure evolution as compared to the pre-emission cathode (Figure 7b). Highly faceted micro-pillars of varying lateral dimension from 2 to 10 μm appeared on the post-emission cathode surface. EDS mapping was performed to study the elemental distribution on the surface of these samples as shown in Figure 7(d-l). EDS mapping generally gives bulk information of the sample (few microns from the sample surface), different from the XPS which is mostly surface sensitive (few nm from the sample surface). The elemental mapping of the as-sintered cathode surface shows nearly uniform distribution of the elements, except the large surface grains, which seem to be Sr- and O- rich and V-deficient, suggesting they are locations where the Sr segregation observed in XPS (Section 4.4) occurs. The EDS mapping of the pre-emission cathode shows uniform distribution of Sr, V and O atoms without any Sr-segregation. The EDS mapping of the post-emission cathode reveals that the micro-pillars have nearly stoichiometric Sr/V ratio and the open region on the cathode surface which is not covered by the micro-pillar is Sr-deficient and V-rich. EDS elemental analysis was also performed at several points on the top surface of these micro-pillars (like Region 1, identified in Figure 7c) and on the open region (like Region 2, identified in Figure 7c) of the cathode surfaces shown in Figure 7c. The point selection for EDS mapping in Region 1 and Region 2 of the emission tested cathode are also shown in Figure S3. The EDS analysis reveals that these micro-pillars (Region 1) have a nearly stoichiometric Sr/V ratio (0.95±0.05). In contrast, the open region, which is not covered by the micro-pillars on the cathode surface (Region 2), is Sr-deficient (Sr/V = 0.43±0.08), again suggesting that there is elemental redistribution during the high temperature cathode activation process. From the nearly stoichiometric Sr/V ratio of the micro-pillars, we speculate these micro-pillars are mostly perovskite $SrVO_3$. On the other hand, the base of these micro-pillars (i.e., Region 2) acts as Sr-supplier and hence gets depleted of Sr. However, further understanding of the surface microstructure evolution and its implication for the $SrVO_3$ cathode is needed to better understand the Sr-segregation process and improve the overall emission performance.

## 4.5. Structural stability assessment of SrVO$_3$

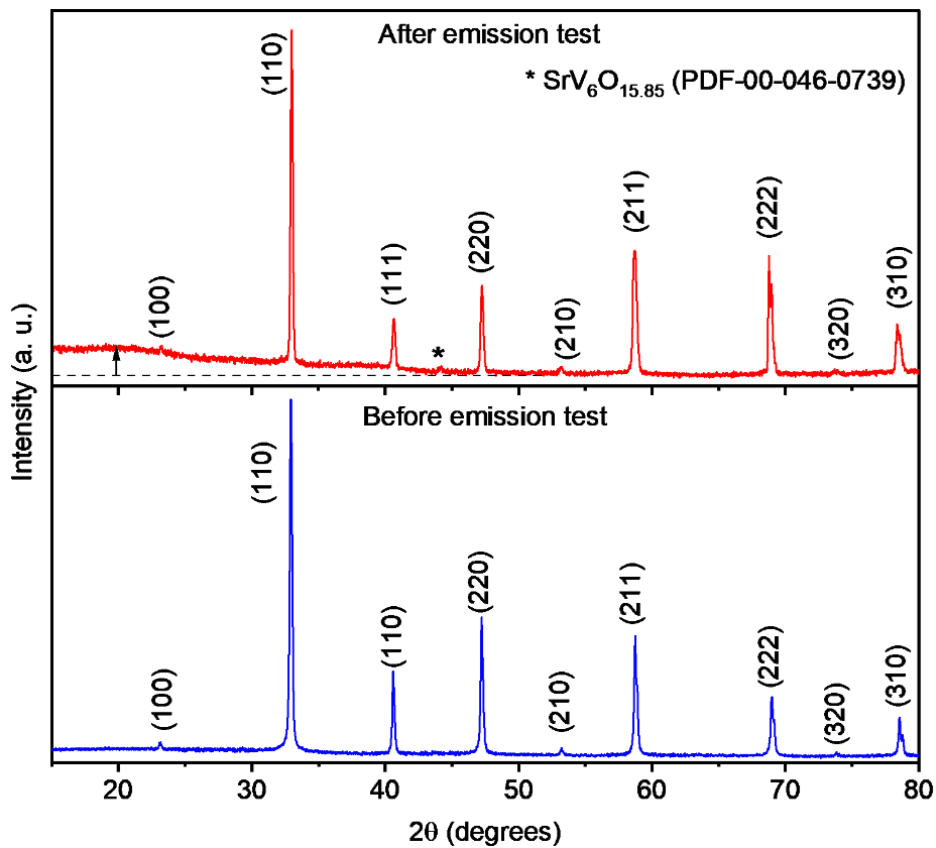

**Figure 8:** XRD pattern of the cathode surface (a) pre-emission cathode and (b) post-emission cathode.

Room temperature XRD measurements of the cathodes were taken after the electron emission test to check the phase stability of the SrVO$_3$ cathode during its operation, which are shown in Figure 8. The XRD pattern of the pre-emission cathode confirms the phase pure perovskite structure of SrVO$_3$. The XRD pattern of the post-emission cathode reveals the weak peak of a non-perovskite V-rich phase (SrV$_6$O$_{15.85}$) and the elevated background intensity at lower angle, suggesting the presence of amorphous phases. XRD analysis also reveals that the position of the most intense XRD peak, (110) slightly moves to the higher angle in the post-emission cathode sample (2Θ ~ 32.98 °) compared to the pre-emission cathode sample (2Θ ~ 32.92 °), which is consistent with Sr cation deficiency (discussed more above in Section 4.4 and 4.5) causing a decrease in lattice parameter. The overall XRD analysis suggests slow chemical and structural evolution of perovskite SrVO$_3$ in ultra-high vacuum and high temperature. The cathode may evaporate Sr from its Sr-segregated surface, resulting in the overall Sr-depletion and evolution of the perovskite structure. As noted above, we currently have no evidence that such chemical and structural evolution will have a negative impact on the electron emission, but further study is warranted to better understand these structural changes and their possible impact on long-term emission behavior in $SrVO_3$.

## 5. Conclusions

Overall, this study finds that $SrVO_3$ polycrystalline perovskite oxide cathodes show significant electron emission with an approximately 2.7-2.8 eV effective work function in the temperature range 1050 to 1250 ºC, has good stability under long term (~ 400 hours) high-temperature and vacuum annealing, under conditions of repeated thermal cycling and continuous pulse electron emission, and can undergo an initial vacuum annealing activation phase likely associated with Sr segregation to the surface. However, some Sr loss and phase transformation is seen over the 400 hours, which may represent a source of long-term degradation. These results generally support that $SrVO_3$ may be a promising cathode candidate for various low power device applications, e.g., electron microscopes.

More specifically, in this work we found high temperature vacuum annealing of $SrVO_3$ cathode helps to gradually enhance its emission current and to realize a modestly low effective work function of about 2.7-2.8 eV. XPS and EDS suggest some parts of the surface have significant Sr and O enrichment under this vacuum annealing, which may create positive Sr-O surface dipoles that decrease the effective work function of $SrVO_3$ cathode during this vacuum activation process. XPS also shows some loss of bulk Sr and XRD shows some small fraction of phase transformed materials after long-time (400 hours) vacuum annealing. These changes led to no observable degradation but suggest possible long-term stability issues with Sr loss and materials degradation. However, more research is required to understand the surface Sr and O segregation phenomena and Sr loss and to fully explore the impact of these factors on operational stability.

Important for realizing practical applications of $SrVO_3$, the activated low effective work function cathode surface shows stable emission performance and is durable during cyclic heating and cooling, suggesting its promising stability, and the cathode shows drift-free steady emission during one hour of continuous pulse electron emission tests at a temperature range from 1135 to 1260 ºC. Despite these promising results, a robust reproducible strategy to reach the lowest reported (approximately 2.3 eV) effective work function value of SrVO$_3$ cathode has not yet been achieved, and reproducibly obtaining this low effective work function may open the door for SrVO$_3$ to be used in higher power applications. Some strategies, such as surface microstructure engineering, incorporation of excess Sr, formulation and manipulation of single crystals, and optimization of the key SrVO$_3$ pellet synthesis parameters of sintering temperature, time and reducing gas condition may be needed to reach this 2.3 eV effective work function reproducibly, and may potentially enable even lower effective work functions. Having reduced effective work function could further lower the operating temperature of a SrVO$_3$ cathode, which in turn may reduce the Sr-evaporation loss from the bulk of the sample. In addition, partially or totally replacing Sr with a lower vapor pressure element may be effective to stabilize the material in ultra-high vacuum and high temperature for even longer time. We believe obtaining robust, reproducible low work function SrVO$_3$ may be achieved through understanding the complex interplay of SrVO$_3$ sample processing and the resulting surface chemistry and chemical processes occurring during activation and emission (e.g., surface Sr-segregation), and the resulting impact on the effective work function.

## Data availability

The thermionic emission data as reported in Figure 2 and 3, and other reported Figure data of the XRD, XPS, EDS results along with their raw data are openly available in figshare at https://doi.org/10.6084/m9.figshare.24352672.v1.

## Author information:

### Corresponding authors

Dane Morgan*

Address: Department of Materials Science and Engineering, University of Wisconsin-Madison, Madison, WI, 53706, USA. Email: ddmorgan@wisc.edu

John Booske*

Address: Department of Electrical Engineering and Computer Science, University of Wisconsin-Madison, Madison, WI, 53706, USA. Email: jhbooske@wisc.edu

### Authors

Md Sariful Sheikh, Ryan Jacobs

Address: Department of Materials Science and Engineering, University of Wisconsin-Madison, Madison, WI, 53706, USA

## Authors contribution

R.J., D.M. and J.B.: Conceptualization, funding acquisition and discussion. M.S.S.: Investigations, discussion and writing (first draft). All authors have reviewed and edited the manuscript.

## Notes:
The authors declare no competing financial interest.

## Acknowledgement
The authors would like to thank the Wisconsin Alumni Research Foundation (WARF) for providing financial research support for this work. The authors gratefully acknowledge the support from Martin Kordesch for providing the heater fixture and high voltage transformer for thermionic electron emission testing.

# Supporting Information for:

# Time-dependence of SrVO$_3$ thermionic electron emission properties

*Md Sariful Sheikh[1], Ryan Jacobs[1], Dane Morgan[1,\*], John Booske[2,\*]*

[1] Department of Materials Science and Engineering University of Wisconsin-Madison, Madison, WI, 53706, USA.
[2] Department of Electrical Engineering and Computer Science University of Wisconsin-Madison, Madison, WI, 53706, USA

Corresponding authors email address: ddmorgan@wisc.edu, jhbooske@wisc.edu

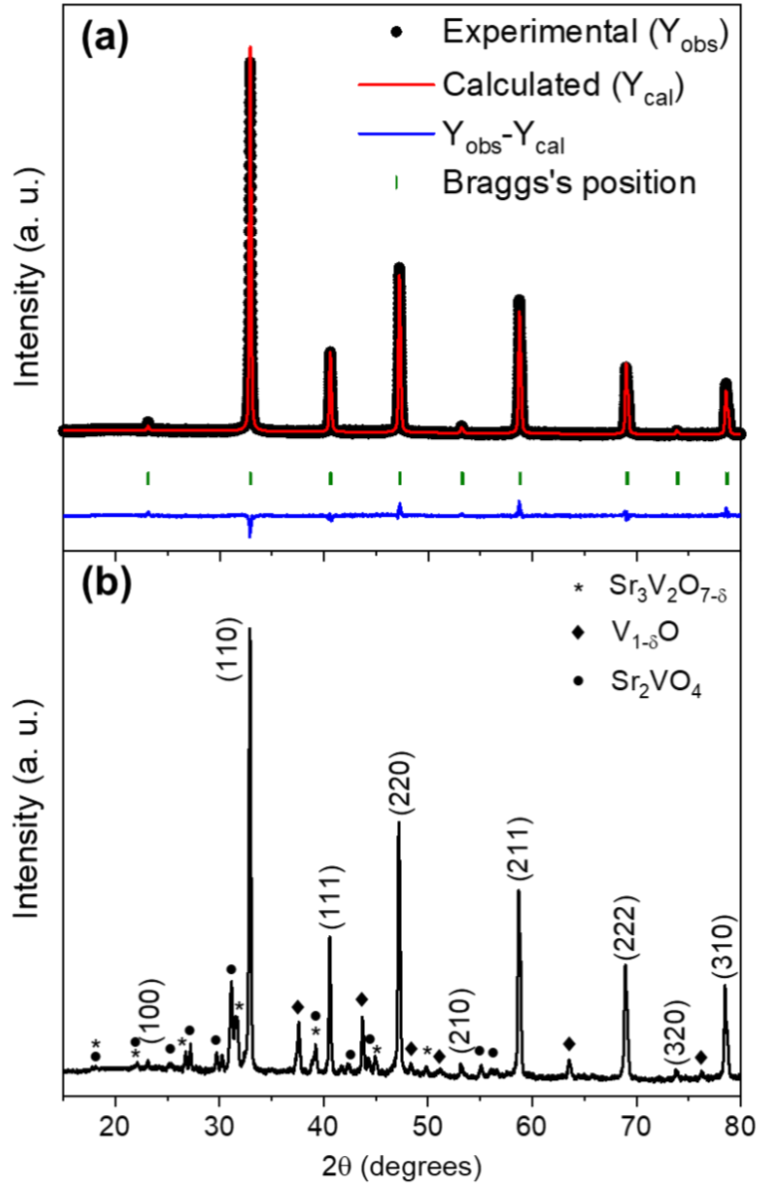

**Figure S1: (a)** X-ray Diffraction (XRD) pattern and Rietveld refinement of the synthesized SrVO$_3$ perovskite oxide powder, **(b)** XRD pattern of as-sintered SrVO$_3$ cathode surface. The as-sintered SrVO$_3$ cathode surface shows the presence of over-reduced non-perovskite phases. Cathode was prepared by scraping the as-sintered cathode surface with a sharp knife. Emission tests were performed on the scraped cathode.

## 1. X-ray Photoemission spectroscopy (V-2p and O-1s core level spectra):

The XPS core-level spectra were fitted using Shirley type background at CasaXPS software. To fit the entire O-1s and V-2p spectra, we used a single Shirley background as reported in previous studies [36]. Separate Shirley background fitting of O-1s and V-2p spectra gives a better fit to the experimental data but underestimates the V content. The separation between the different oxidation peaks and their FWHM values are compared with reported results and are consistent with the previous reported XPS data [31, 35, 36]. The V $2p_{1/2}$ components are fitted within the range from 520 to 525 eV and the V $2p_{3/2}$ peaks are fitted within 512 to 520 eV range. During fitting V $2p_{1/2}$ components a splitting orbital energy gap of 7.5 eV and an aerial ratio of 1:2 is maintained with its V $2p_{3/2}$ components. The FWHMs of the V $2p_{1/2}$ components are within the 3 to 3.5 eV range. A summary of the fitting parameters is represented in Table S2. The deconvolution of V $2p_{3/2}$ peak of the pre-emission test cathode from 512 to 520 eV generates four V peaks with gradually decreasing FWHM in the range from 2 to 1.7 eV, which are shown in Figure S2(a). The peaks at 513.9, 515.4, 516.9, 518.4 eV are associated with $V^{3+}$ hydroxide components, $V^{3+}$ oxide, $V^{4+}$ and $V^{5+}$ oxides respectively. Ideally, V should have $V^{4+}$ oxidation state in $SrVO_3$. But, reduced oxidation states arise due to oxygen vacancy formation during the synthesis process in reducing gas and $V^{5+}$ state arises due to its over oxidation and air exposure. The areal percentages of $V^{3+}$ (hydroxide), $V^{3+}$ (oxide), $V^{4+}$ and $V^{5+}$ states in the cathode before emission test were 10.4, 34.8, 33.5 and 21.3 % of total surface vanadium, respectively. On the other hand, the percentages of $V^{3+}$ (hydroxide), $V^{3+}$ (oxide), $V^{4+}$ and $V^{5+}$ states in post-emission cathode before $Ar^+$ sputtering are 11.6, 30.9, 35.7 and 23.6 %, respectively, as shown in Figure S2(b). The slightly increased percentage of the higher oxidation V states on the post-emission cathode before $Ar^+$ ion sputter surface as compared to pre-emission cathode suggests the surface oxygen enrichment during its vacuum activation process. However, $Ar^+$ ion sputtered post-emission cathode surface reveals the decreased percentage of the higher oxidation V states in the bulk of the cathode as shown in Figure S2(c). The estimated percentages of the $V^{2+}$, $V^{3+}$ (hydroxide), $V^{3+}$ (oxide), $V^{4+}$ and $V^{5+}$ states in post-emission cathode after $Ar^+$ sputtering are 17.66, 4.6, 40.6, 26.1 and 11.1 %, respectively. The percentage reduction of $V^{4+}$ and $V^{5+}$ states and appearance of $V^{2+}$ states in the bulk of $Ar^+$ sputtered post-emission cathode as compared to pre-emission cathode suggest the increased oxygen vacancy within the bulk of the cathode. Deconvolution of the O-1s peaks reveals the presence of lattice oxygen, surface SrO and/or metal oxide hydroxide, absorbed organic species in all the three samples. The position of lattice O-1s peak in the post-emission cathode before $Ar^+$ sputtering (529.6 eV) is shifted towards the lower energy as compared to the pre-emission cathode (530.7 eV) and $Ar^+$ sputtered post-emission cathode (530.8 eV) which is also consistent with surface O enrichment during the vacuum activation of the SrVO$_3$ electron emission cathode.

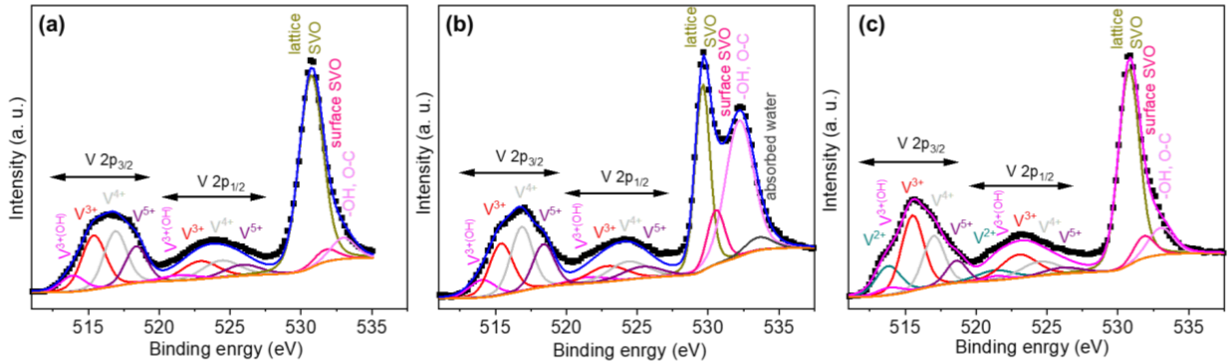

**Figure S2:** Core level V-2p and O-1s atomic spectra of (a) pre-emission cathode (scraped as-sintered cathode), (b) post-emission cathode before Ar$^+$ sputtering, and (c) post-emission cathode after Ar$^+$ sputtering. Black squares and solid lines represent the experimental and fitted data, respectively.

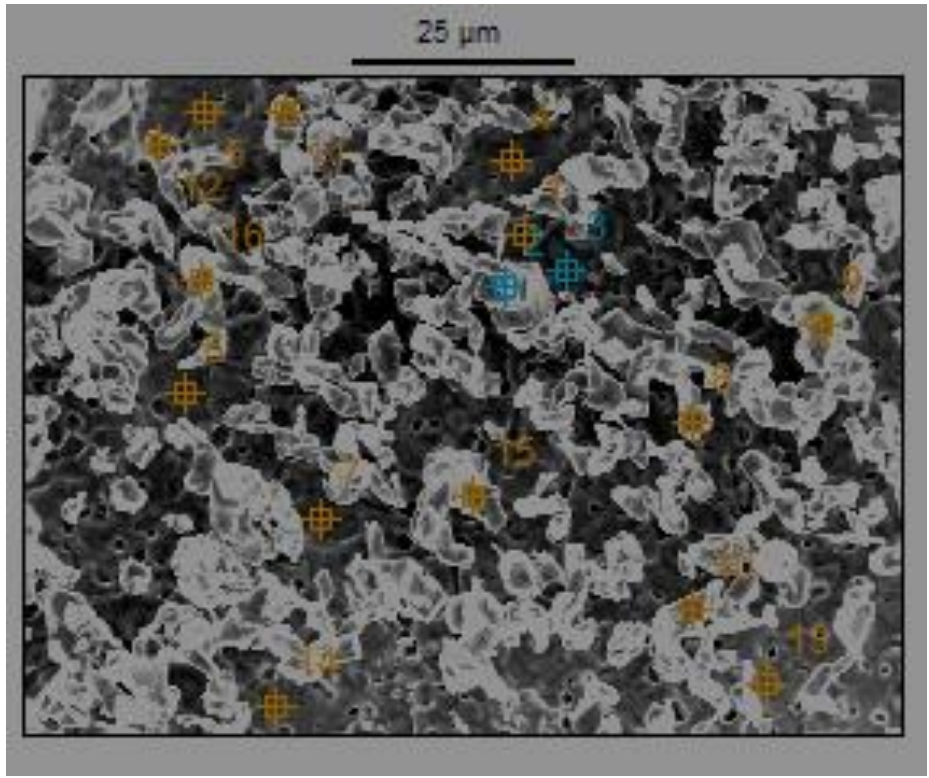

**Figure S3:** EDS point mapping on the surface of the post-emission cathode. Various points on the surface of the cathode were selected randomly. The Sr/V ratio on the micro-pillars and on the open

region on the cathode surface were separately determined from the average of individual ratio obtained from each point of two regions.

**Table S1:** Refined lattice parameters of the sol-gel synthesized SrVO$_3$ perovskite oxide powder.

| Lattice parameters | Reported lattice parameters. (PDF: 04-006-0739) | Atomic positions | Refinement quality parameters |
|---|---|---|---|
| Space group: Pm$\bar{3}$m<br>a = b = c = 3.847(8) Å<br>α = β = γ = 90° | Space group: Pm$\bar{3}$m<br>a = b = c = 3.842 Å<br>α = β = γ = 90° | Sr (0, 0, 0)<br>V (0.5, 0.5, 0.5)<br>O (0.5, 0, 0.5) | $\chi^2$ = 0.341<br>R$_P$ = 17.92, R$_{WP}$ = 12.89, R$_{exp}$ = 22.08 |

**Table S2:** XPS fitting parameters.

| Atomic states | | Binding energy (eV) | | |
|---|---|---|---|---|
| | | Pre-emission cathode (scraped as sintered cathode) | Post-emission cathode before Ar$^+$ sputtering | Post-emission cathode after Ar$^+$ sputtering |
| Sr3d$_{5/2}$ | Sr$^{2+}$ (lattice) | 134.0 | 133.0 | 134.1 |
| V2p$_{3/2}$ | V$^{2+}$ | -- | -- | 513.8 |
| | V$^{3+}$ (hydroxide) | 513.9 | 514.1 | 513.9 |
| | V$^{3+}$ (oxide) | 515.4 | 515.4 | 515.5 |
| | V$^{4+}$ | 516.9 | 516.9 | 517 |
| | V$^{5+}$ | 518.4 | 518.4 | 518.6 |
| O1s$_{1/2}$ | O$^{2-}$ (lattice) | 530.7 | 529.6 | 530.8 |
| | O$^{2-}$ (surface) | 531.7 | 530.6 | 531.8 |
| | O$^{2-}$ (-OH, -O-C) | 533 | 532.2 | 533 |
| | O$^{2-}$ (absorbed H$_2$O) | -- | 533.6 | -- |

## References:

[1] Jenkins, R. O. A review of thermionic cathodes. *Vacuum* **1969**, 19 (8), 353-359. https://doi.org/10.1016/S0042-207X(69)80077-1

[2] Gao, J. Y.; Yang, Y. F.; Zhang, X. K.; Li, S. L.; Wang, J. S. A review on recent progress of thermionic cathode. *Tungsten* **2020**, 2, 289-300. https://doi.org/10.1007/s42864-020-00059-1

[3] Gaertner, G.; Engelsen, D. den. Hundred years anniversary of the oxide cathode-A historical review. *Appl. Surf. Sci.* **2005**, 251, 24-30. https://doi.org/10.1016/j.apsusc.2005.03.214

[4] Zhang, P.; Ang, Y. S.; Garner, A. L.; Valfells, Á.; Luginsland, J. W.; Ang L. K. Space-charge limited current in nanodiodes: Ballistic, collisional, and dynamical effects. *J. Appl. Phys.* **2021**, 129 (10), 100902. https://doi.org/10.1063/5.0042355

[5] Booske, J. H. Plasma physics and related challenges of millimeter-wave-to-terahertz and high power microwave generation. *Phys. Plasmas* **2008**, 15, 055502. https://doi.org/10.1063/1.2838240

[6] Gilmour, A. S. Klystrons, traveling wave tubes, magnetrons, crossed-field amplifiers, and gyrotrons. Artech House, Norwood, MA, 2011.

[7] Jin, F.; Beaver, A. High thermionic emission from barium strontium oxide functionalized carbon nanotubes thin film surface. *Appl. Phys. Lett.* **2017**, 110 (21), 213109. http://dx.doi.org/10.1063/1.4984216

[8] Singh, A. K.; Shukla, S. K.; Ravi, M.; Barik, R. K. A Review of Electron Emitters for High-Power and High-Frequency Vacuum Electron Devices, *IEEE Trans. Plasma Sci.* **2020**, 48 (10), 3446-3454. https://doi.org/10.1109/TPS.2020.3011285

[9] Schwede, J.; Bargatin, I.; Riley, D. C.; Hardin, B. E.; Rosenthal, S. J.; Sun, Y.; Schmitt, F.; Pianetta, P.; Howe, R. T.; Shen, Z.-X.; Melosh, N. A. Photon-enhanced thermionic emission for solar concentrator systems. *Nature Mater.* **2010**, 9, 762-767. https://doi.org/10.1038/nmat2814

[10] Nirantar, S.; Ahmed, T.; Bhaskaran, M.; Han, J.-W.; Walia, S.; Sriram, S. Electron emission devices for energy-efficient systems. *Adv. Intell. Syst.* **2019**, 1 (4), 1900039. https://doi.org/10.1002/aisy.201900039

[11] Jin, F.; Miruko, A.; Litt, D.; Zhou, K. Functionalized carbon nanotubes for thermionic emission and cooling applications. *J. Vac. Sci. Technol. A* **2022**, 40 (1), 013415. https://doi.org/10.1116/6.0001467

[12] Markevich, A.; Hudak, B. M.; Madsen, J.; Song, J.; Snijders, P. C.; Lupini, A. R.; Susi, T. Mechanism of Electron-Beam Manipulation of Single-Dopant Atoms in Silicon. *J. Phys. Chem. C* **2021**, 125 (29), 16041-16048. https://doi.org/10.1021/acs.jpcc.1c03549

[13] Datas, A.; R. Vaillon. Thermionic-enhanced near-field thermophotovoltaics. *Nano Energy* **2019**, 61, 10-17. https://doi.org/10.1016/j.nanoen.2019.04.039

[14] Holste, K.; Dietz, P.; Scharmann, S.; Keil, K.; Henning, T.; Zschätzsch, D.; Reitemeyer, M.; Nauschütt, B.; Kiefer, F.; Kunze, F.; Zorn, J.; Heiliger, C.; Joshi, N.; Probst, U.; Thüringer, R.; Volkmar, C.; Packan, D.; Peterschmitt, S.; Brinkmann, K. -T.; Zaunick, H.-G.; Thoma, M. H.; Kretschmer, M.; Leiter, H. J.; Schippers, S.; Hannemann, K.; Klar, P. J. Ion thrusters for electric propulsion: Scientific issues developing a niche technology into a game changer. *Rev. Sci. Instrum.* **2020**, 91 (6), 061101. https://doi.org/10.1063/5.0010134

[15] Lin, L.; Jacobs, R.; Ma, T.; Chen, D.; Booske, J.; Morgan, D. Work function: fundamentals, measurement, calculation, engineering, and applications. *Phys. Rev. Appl.* **2023**, 19 (3), 037001. https://doi.org/10.1103/PhysRevApplied.19.037001

[16] Seif, M. N.; Zhou, Q.; Liu, X.; Balk, T. J.; Beck, M. J. Sc-containing (scandate) thermionic cathodes: mechanisms for Sc enhancement of emission. *IEEE Trans. Electron Devices* **2022**, 69 (7), 3523-3534. https://doi.org/10.1109/TED.2022.3172054

[17] Hasan, M. M.; Kisi, E.; Sugo, H. Structural and thermionic emission investigations of perovskite $BaHfO_3$ based low work function emitters. *Mater. Sci. Eng. B* **2023**, 296, 116679. https://doi.org/10.1016/j.mseb.2023.116679

[18] Kimura, S.; Yoshida, H.; Miyazaki, H.; Fujimoto, T.; Ogino, Akihisa. Enhancement of thermionic emission and conversion characteristics using polarization- and band-engineered n-type AlGaN/GaN cathodes. *J. Vac. Sci. Technol. B* **2021**, 39, 062207. https://doi.org/10.1116/6.0001357

[19] Mondal, S.; Rau, A. V.; Lu, K.; Li, J.-F.; Viehland, D. Multicomponent hexaborides with low work functions by ultra-fast high temperature sintering. *Open Ceram.* **2023**, 16, 100479. https://doi.org/10.1016/j.oceram.2023.100479

[20] Ma, T.; Jacobs, R.; Booske, J.; Morgan, D. Work function trends and new low-work-function boride and nitride materials for electron emission applications. *J. Phys. Chem. C* **2021**, 125 (31), 17400-17410. https://doi.org/10.1021/acs.jpcc.1c04289

[21] Schmidt, P. H.; Longinotti, L. D.; Joy, D. C.; Ferris, S. D.; Leamy, H. J.; Fisk, Z. Design and optimization of directly heated $LaB_6$ cathode assemblies for electron-beam instruments. *J. Vac. Sci. Technol.* **1978**, 15 (4), 1554-1560. https://doi.org/10.1116/1.569786

[22] Taran, A.; Voronovich, D.; Plankovskyy, S.; Paderno, V.; Filipov, V. Review of $LaB_6$, Re-W dispenser, and $BaHfO_3$-W cathode development. *IEEE Trans. Electron Devices* **2009**, 56 (5), 812-817. https://doi.org/10.1109/TED.2009.2015615

[23] Pedrini, D.; Albertoni, R.; Paganucci, F.; Andrenucci, M. Modeling of $LaB_6$ hollow cathode performance and lifetime. *Acta Astronaut.* **2015**, 106, 170-178. https://doi.org/10.1016/j.actaastro.2014.10.033

[24] Kirkwood D. M.; Gross, S. J.; Balk, T. J.; Beck, M. J.; Booske, J.; Busbaher, D.; Jacobs, R.; Kordesch, M. E.; Mitsdarffer, B.; Morgan, D.; Palmer, W. D.; Vancil, B.; Yater, J. E. Frontiers in thermionic cathode research. *IEEE Trans. Electron Devices* **2018**, 65 (6), 2061-2071. https://doi.org/10.1109/TED.2018.2804484

[25] Swartzentruber, P. D.; Balk, T. J.; Effgen, M. P. Correlation between microstructure and thermionic electron emission from Os-Ru thin films on dispenser cathodes. *J. Vac. Sci. Technol. A* **2014**, 32 (4), 040601. http://dx.doi.org/10.1116/1.4876337

[26] Jacobs, R.; Booske, J.; Morgan, D. Understanding and controlling the work function of perovskite oxides using density functional theory. *Adv. Funct. Mater.* **2016**, 26 (30), 5471-5482. https://doi.org/10.1002/adfm.201600243

[27] Ma, T.; Jacobs, R.; Booske, J.; Morgan, D. Discovery and engineering of low work function perovskite materials. J. Mater. Chem. C **2021**, (9), 12778-12790. https://doi.org/10.1039/D1TC01286J

[28] Lin, L.; Jacobs, R.; Chen, D.; Vlahos, V.; Lu-Steffes, O.; Alonso, J. A.; Morgan, D.; Booske, J. Demonstration of low work function perovskite SrVO$_3$ using thermionic electron emission. *Adv. Funct. Mater.* **2022**, 32 (41), 2203703. https://doi.org/10.1002/adfm.202203703

[29] Lin, L.; Jacobs, R.; Morgan, D.; J. Booske, J. Investigating thermionic emission properties of polycrystalline perovskite BaMoO$_3$. *IEEE Trans. Electron Devices*, **2023**, 70 (4),1871-1877. https://doi.org/10.1109/TED.2023.3242936

[30] Sheikh, M. S.; Lin, L.; Jacobs, R.; Morgan, D.; Booske, J. SrVO$_3$ electron emission cathodes with stable, >250 mA/cm$^2$ current density. 24$^{th}$ International Vacuum Electronics Conference (IVEC), Chengdu, China, **2023**, pp. 1-2. https://doi.org/10.1109/IVEC56627.2023.10157940

[31] Bourlier, Y.; Frégnaux, M.; Bérini, B.; Fouchet, A.; Dumont, Y.; Aureau, D. XPS monitoring of SrVO$_3$ thin films from demixing to air ageing: The asset of treatment in water. *Appl. Surf. Sci.* **2021**, 553, 149536. https://doi.org/10.1016/j.apsusc.2021.149536

[32] Koo, B.; Kim, K.; Kim, J. K.; Kwon, H.; Han, J. W.; Jung, W. Sr segregation in perovskite oxides: why it happens and how it exists. *Joule* **2018**, 2 (8), 1476-1499. https://doi.org/10.1016/j.joule.2018.07.016

[33] Ostrovskiy, E.; Huang, Y.-L.; Wachsman, E. D. Effects of surface chemical potentials on cation segregation. *J. Mater. Chem. A* **2021**, 9, 1593-1602. https://doi.org/10.1039/D0TA08850A

[34] Choi, M.; Ibrahim, I. A. M.; Kim, K.; Koo, J. Y.; Kim, S. J.; Son, J.-W.; Han, J. W.; Lee, W. Engineering of charged defects at perovskite oxide surfaces for exceptionally stable solid oxide fuel cell electrodes. *ACS Appl. Mater. Interfaces* **2020**, 12 (19), 21494-21504. https://doi.org/10.1021/acsami.9b21919

[35] Bachelet, R.; Sánchez, F.; Palomares, F. J.; Ocal, C.; Fontcuberta, J. Atomically flat SrO-terminated SrTiO$_3$ (001) substrate. *Appl. Phys. Lett.* **2009**, 95, 141915. https://doi.org/10.1063/1.3240869

[36] Silversmit, G.; Depla, D.; Poelman, H.; Marin, G. B.; Gryse, R. D. Determination of the V2p XPS binding energies for different vanadium oxidation states (V$^{5+}$ to V$^{0+}$). *J. Electron. Spectros. Relat. Phenomena* **2004**, 135, 167-175. https://doi.org/10.1016/j.elspec.2004.03.004